\documentclass[twocolumn,epjc3]{svjour3}          

\RequirePackage[T1]{fontenc}

\smartqed  

\RequirePackage{graphicx}
\RequirePackage{mathptmx}      
\RequirePackage{flushend}
\RequirePackage[numbers,sort&compress]{natbib}
\RequirePackage[colorlinks,citecolor=blue,urlcolor=blue,linkcolor=blue]{hyperref}

\journalname{International Journal of Information Security}

\usepackage{glossaries}
\usepackage{xspace}
\usepackage{booktabs}
\usepackage{fontawesome5}
\usepackage{siunitx}
\usepackage{tabularray}
\UseTblrLibrary{booktabs}

\AtBeginDocument{%
}

\newcommand{\apk}{{APK}\xspace}
\newcommand{\apks}{{APKs}\xspace}
\newcommand{\ie}{i.e.,\xspace}
\newcommand{\eg}{e.g.,\xspace}

\newcommand{\etal}{{et al.}\xspace}
\newcommand{\soa}{{state-of-the-art}\xspace}

\newcommand{\myparagraph}[1]{\noindent\textbf{#1.}}

\newcommand{\todo}[1]{}
\renewcommand{\todo}[1]{{\color{red} TODO: {#1}}}

\newcommand{\yes}{\scalebox{0.7}{\faCheck}\xspace}
\newcommand{\no}{{-}\xspace}
\newcommand{\static}{static\xspace}
\newcommand{\dynamic}{dynamic\xspace}
\newcommand{\centralized}{\texttt{C}\xspace}
\newcommand{\fl}{\texttt{FL}\xspace}
\newcommand{\high}{\texttt{H}\xspace}
\newcommand{\medium}{\texttt{M}\xspace}
\newcommand{\low}{\texttt{L}\xspace}
\newacronym{gdpr}{GDPR}{General Data Protection Regulation}
\newacronym{ml}{ML}{Machine Learning}
\newacronym{dl}{DL}{Deep Learning}
\newacronym{os}{OS}{Operating System}
\newacronym{pha}{PHA}{Potentially Harmful Application}
\newacronym{vt}{VT}{VirusTotal}
\newacronym{fl}{FL}{Federated Learning}
\newacronym{PII}{PII}{Personal Identifiable Information}
\hypersetup{
    urlcolor=black
}

\authorrunning{Massidda et al.}

\begin{document}

\title{Don't Trust Us: A privacy-by-design android malware detection pipeline}




\author{Emmanuele Massidda$^{1,*}$ \and
Diego Soi$^{1}$ \and
Giorgio Giacinto$^{1,2}$
}

\institute{
$^{*}$ Corresponding Author
\\
$^{1}$ Department of Electrical and Electronic Engineering (DIEE),
University of Cagliari, 09123 Cagliari, Italy
\\
\email{emmanuele.massidda@unica.it, diego.soi@unica.it, giorgio.giacinto@unica.it}
\\
$^{2}$ Consorzio Interuniversitario Nazionale per l’Informatica (CINI), Rome, Italy
}

\date{Received: date / Accepted: date}

\maketitle

\begin{abstract}
Android malware detection increasingly relies on collecting and processing sensitive user data, including device identifiers, network artifacts, and runtime traces, while privacy is too often treated as a secondary concern. Existing privacy-aware approaches typically enforce privacy after data collection, for example, through anonymization, encryption, or federated learning, yet still require access to user information and therefore demand a high level of user trust in systems that already operate with privileged access to device activity. We argue that this requirement should be removed rather than managed. Android malware detection should be privacy-aware by design, so that effective analysis does not depend on sensitive data being accessed in the first place. To this end, we first formalize a set of design requirements for privacy-by-design detection and then implement each requirement in a comprehensive pipeline. First, static analysis is performed to extract relevant data from each APK, following the Drebin representation, which is then submitted to an SVM after vectorization. The model is equipped with a dual-reject threshold rule that either commits to a confident decision or defers uncertain samples to a dynamic analysis stage within a sandboxed environment, so that genuine user information never enters the analysis loop.
Results confirm that, on a temporally split dataset spanning from 2024 to 2025, the pipeline achieves an F\textsubscript{1} score of 0.87 with the first static analysis stage, deferring only 6.7\% of test samples to secondary dynamic analysis. Additionally, dynamic sandboxing helps recognize applications' maliciousness with high confidence without extracting any sensitive data. These results demonstrate that strong detection performance is achievable without sacrificing user privacy.
\end{abstract}


\section{Introduction}
Among mobile \gls{os}, Android has become the most widely employed due to its versatility and its ability to adapt to diverse hardware platforms and deployment environments. As a consequence, Android has become a primary target for malicious software (malware) that threatens the confidentiality, integrity, and availability of the data stored on user devices. 
In response, the research community and industry have developed numerous Android malware detection techniques, ranging from signature- and heuristic-based approaches to modern machine-learning systems~\cite{DREBIN, Mariconti_MamaDroid, Zhang_APIGraph, Soi_FCG_Explainability_2024, Cui2023_SE, Spreitzenbarth2015_IS} that leverage static, dynamic, and hybrid analyses. In recent years, data-driven detection has become especially dominant: richer telemetry and larger datasets typically translate into higher accuracy and faster adaptation to new malware variants (e.g., zero-day attacks). However, this accuracy is often achieved by collecting and processing large volumes of sensitive information, including application inventories, runtime traces, system logs, and network traffic artifacts~\cite{Li_Zhai_Zhang_Quan_2014,Norouzian2021_ISC}. The core conflict is structural and mainly depends on the fact that effective malware detection depends on more data, while privacy protection, even in relation to GDPR~\cite{gdpr} and privacy-aware regulations, demands strict minimization, purpose limitations and reduced data collection of personal data, also called~\gls{PII}. A typical response to this conflict is to treat privacy as a data-handling problem, \ie once data is collected, it can be “secured” through access controls, encryption, anonymization, or policy rules~\cite{Anonymisation_2018}. Yet in the context of mobile security systems, the mere availability of sensitive telemetry to an anti-malware pipeline already creates privacy risk. In other words, privacy is not only about how data is managed, but whether the detection process needs to touch sensitive data at all. This is particularly relevant on Android, where detection pipelines may be deployed at scale (\eg in cloud-assisted services, vendor telemetry, or third-party security apps) and thus can enable aggregation, user profiling, and secondary use~\cite{Reardon2019_USENIX}.

This concern is reinforced when considering the types of signals used by conventional detection pipelines. Static analysis is frequently characterized as relatively less privacy-invasive because it can be performed without directly observing user behavior. However, privacy risks arise in practice when static results are uploaded to centralized servers, linked to persistent identifiers, or combined with additional device context. Even simple artifacts such as the list of installed applications or the presence of specific apps can support sensitive inferences about personal habits. Dynamic and network-based approaches introduce further challenges because they inherently observe behavior. For instance, network-monitoring techniques may rely on traffic analysis and packet inspection, therefore exposing sensitive information such as communication endpoints and browsing patterns. 
These observations do not imply that dynamic artifacts are unusable, but they highlight that many practical detection strategies implicitly require levels of visibility that are misaligned with privacy minimization.

At a time when digital services increasingly prioritize efficiency and functionality, privacy is often treated as a secondary concern. In this context, users must be informed about how their data is collected, used, and processed, especially when security tools operate with privileged access to personal devices. Systems such as anti-malware solutions are expected to protect users, yet their operation often requires full visibility into device activity and application behavior~\cite{avast_permissions}. For this reason, transparency in how these systems operate and what data they rely on is critical. This paper, therefore, advances the position that Android malware detection should be privacy-aware by design. Privacy requirements should be integrated directly into the detection methodology so that effective analysis does not depend on collecting sensitive user data. 
Building on this position, we present a privacy-aware pipeline for Android malware detection that aims to preserve privacy throughout the analysis. Rather than asking users to trust that the detector will handle sensitive data responsibly, a privacy-by-design pipeline removes that trust requirement by ensuring sensitive data is never accessed in the first place, not through assurances about how it is managed.
Specifically, we propose a four-stage pipeline that pairs static extraction with a dual-threshold classifier, deferring only uncertain samples to a dynamic analysis stage operating on synthetic data. Our contributions are: (i) a set of design requirements for privacy-by-design Android malware detection; (ii) a concrete pipeline that instantiates such requirements; and (iii) an empirical evaluation of the system.
The remainder of the paper is organized as follows. \autoref{sec:background} introduces background on Android applications and privacy threats in mobile security systems, while \autoref{sec:sota} reviews privacy risks in conventional Android malware detection in the context of current literature approaches, clarifying their limitations. \autoref{sec:meth} presents the proposed privacy-aware detection pipeline and its design requirements. Eventually, \autoref{sec:eval} presents the experimental settings and results, and~\autoref{sec:conclusion} concludes the paper discussing limitations and future works.

\section{Background}
\label{sec:background}

This section discusses background information about Android \apks in~\autoref{subsec:background:apk}, and the privacy risks of malware detectors in~\autoref{subsec:background:risks}.

\subsection{Android APKs}\label{subsec:background:apk}

Android \apk is the format with which Android applications are distributed, which follows a structure containing different resources. Indeed, it consists of \emph{(i)} lib directory holding native libraries for each compilation platform (\eg x86, and ARM64); \emph{(ii)} res and assets directories with compiled and not compiled resources (\eg images, videos, and layouts); \emph{(iii)} the AndroidManifest.xml that describes essential information for the OS and the \apk builder like requested permissions, app modules and intent filters\footnote{App Modules and Intent filters represent the entry points of the application subdivided into activities, services, receivers and providers, and the actions that can spawn those entry points, respectively.}, and hardware features, \ie the hardware required by the application; and eventually \emph{(iv)} one or more DEX files that holds the executable code obtaining from the compilation of Java/Kotlin source code, and executed by ART (Android RunTime) that translates DEX code into ARM to speed execution.

\subsection{Privacy Risks in Android Malware Detection}\label{subsec:background:risks}

In Android malware detection systems, privacy risks arise at various stages of the detection pipeline, including not only \gls{ml} and \gls{dl} approaches, but also signature-based techniques. In the following, we identify and briefly outline the main categories of risks.

\myparagraph{Data Collection} It regards the risk associated with the privacy of personal data from misuse or unauthorized access. Indeed, according to EU \gls{gdpr}~\cite{gdpr}, personal data (\eg names, location, and biometric data) should be protected, as, when used alone and not properly anonymized, they may uniquely identify an individual.

\myparagraph{Feature Extraction} It concerns the risk associated with the feature extraction phase of an antimalware technique. Typically, all techniques require an extraction of features from applications. However, the \gls{ml} and \gls{dl} pipelines tend to employ a large number of features that represent the application with suitable, structured characteristics, to be fed to learning models that aim to learn correlations in the data. Although this phase is technically privacy-aware, as it abstracts raw inputs, it may still pose privacy risks because features may still encode sensitive data or behavioral patterns (\eg network byte packets).

\myparagraph{Training and Inference} It regards the risk associated with the fact that an attacker can deduce sensitive information about a user or the training data by simply interacting with a deployed learning model. Examples of these attacks are \emph{membership inference}~\cite{Hu2021_ACM} when the attacker queries the model to understand if specific data are included in the training data, and \emph{model inversion}~\cite{Zhou2025_arxiv} when the attacker can reverse engineer the model to extract information about it. In this work, we do not consider it when designing a privacy-aware detection pipeline, as these issues are not directly related to the privacy risks arising from the data collection and processing stages addressed by our framework.

\section{Related Work}
\label{sec:sota} 
In this section, we review techniques and methodologies that, while effective at detecting malicious applications, inherently depend on the collection and processing of sensitive data, and highlight those that fail to account for user and data privacy.~\autoref{tab:detect_comparison} summarizes Android malware detectors and their corresponding privacy risks. 
Given the extensive literature in this area, we selected a representative subset of recent works that explicitly address privacy-preserving malware detection from different perspectives. The goal is not to provide a comprehensive survey, but rather to highlight different strategies and trade-offs adopted to balance detection effectiveness and privacy preservation.

In particular, we categorize approaches based on \emph{Analysis} and \emph{Approach}, defining the extraction phase type (\ie static or dynamic), and the training approach (\ie centralized or \gls{fl} based). Additionally, we define \emph{Overhead}, which represents the complexity of the approach in terms of the overhead of both feature extraction and the training approach. Typically, we can state that \gls{fl} methods have higher overhead than centralized approaches due to the complexity of extracting and training models within constrained devices.
Then, we classify the approaches based on extracted data, and their privacy risks are subdivided into the categories as outlined in~\autoref{subsec:background:risks}. As is noticeable, selected dynamic approaches suffer from all privacy risks, as sensitive data can be identified in each phase of the \gls{ml} pipelines, while static approaches typically do not include \emph{data} and \emph{feature} risks, while \emph{train} and \emph{inference} are still present like any other machine learning approach.

In this section, we highlight which traditional detection techniques fail to account for user privacy (\autoref{subsec:sota:not_aware}), and those that take into account privacy risks, and try to counteract them (\autoref{subsec:sota:aware})

\begin{table*}[t]
\caption{Comparison of Android malware detectors privacy not-aware and privacy-aware based on \emph{Analysis}, \emph{Extracted features}, \emph{Approach} (\centralized centralized, \fl Federated-learning, \texttt{S} signature-based), \emph{Overhead} (\low low, \medium medium, \high high), and \emph{privacy-aware} and \emph{Privacy-risks} (\yes yes, \no no).}
\label{tab:detect_comparison}
\centering
\begin{tblr}{
  colspec = {p{3.2cm} c c c c c ccc},
  rows = {rowsep = 3pt, valign = m},
  hline{1,Z} = {solid, wd=1pt},
  hline{2} = {7-10}{solid, leftpos=-0.5, rightpos=-0.5, endpos},
  hline{3,4,6,9,10,11,14} = {solid},
  vline{2-7} = {solid}
}
& & & & & & \SetCell[c=3]{c}{\textbf{Privacy Risks}}\\
\textbf{Detector} & \textbf{Analysis} & 
\textbf{Data} & 
\rotatebox{90}{\textbf{Approach}} & 
\rotatebox{90}{\textbf{Overhead}} &
\rotatebox{90}{\textbf{Privacy-aware}} &
\rotatebox{90}{Data Collection} &
\rotatebox{90}{Feature Extract.} &
\rotatebox{90}{Train \& Inference}\\

\hline

Bhat \etal~\cite{BHAT2023103277} 
& \dynamic & 
system calls &
\centralized & \medium &
\no & \yes & \yes & \yes\\

\SetCell[r=2]{l}{Chen \etal~\cite{chen_ransomware_realtime}} &
\SetCell[r=2]{c} \dynamic & 
system calls & 
\SetCell[r=2]{c} \centralized & \SetCell[r=2]{c} \medium &
\SetCell[r=2]{c} \no &
\SetCell[r=2]{c} \yes &
\SetCell[r=2]{c} \yes &
\SetCell[r=2]{c} \yes \\
& & UI monitoring & & & & & & \\

\SetCell[r=2]{l}{Norouzian \etal~\cite{Norouzian2021_ISC}} &
\SetCell[r=2]{c} {\static\\\dynamic} & 
network flows & 
\SetCell[r=2]{c} \centralized & \SetCell[r=2]{c} \medium &
\SetCell[r=2]{c} \no &
\SetCell[r=2]{c} \yes &
\SetCell[r=2]{c} \yes &
\SetCell[r=2]{c} \yes \\
& & opcode & & & & & & \\

\midrule\midrule

Hsu \etal~\cite{Hsu_Wang_Fan_Sun_Ban_Takahashi_Wu_Kao_2020} & \static & API calls & 
\fl & \high & \yes & \no & \no & \yes\\

{Lee \etal~\cite{Lee_2023}\\Ciaramella \etal~\cite{Ciaramella2026_IST}} & \static & 
\apk bytecode &
\fl & \high & \yes & \no & \no & \yes\\

APIGraph~\cite{Zhang_APIGraph} &
\static & 
API calls & 
\centralized & \low &
\yes & \no & \no & \yes \\

\SetCell[r=3]{l,m} DREBIN~\cite{DREBIN} &
\SetCell[r=3]{c,m} \static & 
manifest info &
\SetCell[r=3]{c,m} \centralized & \SetCell[r=3]{c,m} \low & 
\SetCell[r=3]{c,m} \yes &
\SetCell[r=3]{c,m} \no &
\SetCell[r=3]{c,m} \no &
\SetCell[r=3]{c,m} \yes \\
& & API calls & & & {} & {} & {} & {} \\
& & URLs & & & {} & {} & {} & {} \\

\SetCell[r=2]{l} AndroMD~\cite{Prasad2025} & 
\SetCell[r=2]{c} \static & 
permissions & 
\SetCell[r=2]{c} \texttt{S} & 
\SetCell[r=2]{c} \low &
\SetCell[r=2]{c} \yes & \SetCell[r=2]{c} \no & 
\SetCell[r=2]{c} \no & 
\SetCell[r=2]{c} \no \\
& & API calls & & & & & & 

\end{tblr}
\end{table*}

\subsection{Privacy not-aware Android Malware Detection}\label{subsec:sota:not_aware}

\myparagraph{Static Analysis-based techniques}
Many traditional anti-malware systems rely on static analysis of the applications, examining application manifest information, its code, and resources without executing the app~\cite{Altaha2024_Access}.
For these reasons, they are generally considered less intrusive from a privacy perspective as they do not typically require access to user data or behavioral traces~\cite{DREBIN, Zhang_APIGraph, Mariconti_MamaDroid}. However, privacy concerns arise when detection pipelines extend static code inspection with auxiliary data collected from the device~\cite{Yao_2020}.

Commercial security applications may introduce \emph{data collection} risks by gathering data for third parties, while the majority collect sensitive information, including personal data and behavioral patterns. In particular, the \emph{data collection} privacy risks increase when these data are uploaded to centralized servers, enabling data aggregation, facilitating user profiling, and exposing sensitive information beyond what is necessary for detection~\cite{Yao_2020,Hu2019_SANER}.
Additionally, security applications may introduce privacy risks by requiring access to Accessibility Services for features like URL scanning and phishing prevention~\cite{avast_permissions}. This grants them broad capabilities to observe the user’s screen and interact with other apps~\cite{android_accessibility_service}, thereby increasing privacy risks due to extensive visibility and control over the device.

Android introduced Google Play Protect~\cite{Play_Protect} as a default security feature integrated with the OS. It aims to detect \gls{pha} through a combination of local scans and cloud scanning prior to app publication on the Google Play Store. Local scans occur daily, while cloud scans utilize static approaches enhanced by machine learning and heuristics. However, Play Protect is not open source, limiting transparency regarding data collection and privacy concerns.

In summary, while static analysis is designed to be privacy-preserving, its effectiveness can vary based on implementation choices (\eg offline or online scanning), scanning methods (\eg large-scale scanning and accessibility services), and the use of machine learning techniques, which may introduce additional privacy risks.

\myparagraph{Dynamic analysis-based techniques}
Dynamic analysis techniques monitor the application as it executes on a sandboxed or a real-world device and typically collect data artifacts that may be traced back to end users, especially when real-time monitoring is used.

Several Android malware detectors leverage system and API call traces as behavioral indicators~\cite{BHAT2023103277}, potentially exposing sensitive information like device identifiers and personal attributes to identify malicious activities related to cryptography or data leakage~\cite{sutter_dynamic}.
Other works, instead, goes beyond system tracing by observing fine-grained user interactions in real-time. Chen~\etal~\cite{chen_ransomware_realtime} detects Android ransomware by examining not only syscall traces but also the widget interfaces rendered during specific activities to determine, for example, whether a file-encryption activity is initiated by the user. When this analysis is done on physical devices, and when data is stored over time, this approach raises significant privacy concerns due to potential user profiling and the misuse of collected data.  
Other approaches monitor network patterns and rely on deep packet inspection, which often involves redirecting an application's traffic through a proxy to circumvent TLS and certificate pinning protections~\cite{Norouzian2021_ISC}. In this case, the privacy concerns are great since sensitive information (\eg domains, IP addresses, and packet content) can be captured. Privacy risks become more evident when unprocessed network data is transmitted to third-party servers, as this may enable re-identification. Moreover, even though these works manage encrypted traffic, or they represent network flow bytes as images~\cite{Aldini2024_ARES, Norouzian2021_ISC} they may still raise concerns as they consider the entire traffic without excluding application level, and because users' activity may be reconstructed by looking at network statistics~\cite{Conti2015_codaspy}. The privacy risks concern the type of analysis performed and the devices used to perform it. In particular, if these security approaches are used to analyze real user applications, the centralized server analyzing network traffic can access sensitive information about users' behavior.

\subsection{Privacy Aware Android Malware Detection}\label{subsec:sota:aware}

\myparagraph{Federated Learning Approaches}
A limited number of studies in \soa explicitly investigate privacy-aware Android malware detection, and often propose \gls{fl} approaches as a suitable design choice. 
Indeed, Federated learning is a distributed machine learning approach that enables decentralized model training without sharing raw data with a centralized server, thereby limiting privacy risks~\cite{Shokry2015_CCS, Google_FL}.

Hsu \etal~\cite{Hsu_Wang_Fan_Sun_Ban_Takahashi_Wu_Kao_2020} address the privacy limitations of previous Android malware detection approaches, proposing a \gls{fl} architecture that trains models locally and aggregates them centrally on a server. To further protect the exchanged information, the aggregation process relies on Secure Multi-Party Computation (SMPC) combined with secret-sharing mechanisms, preventing the server or other parties from reconstructing individual model updates. Their evaluation demonstrates that the performance is comparable to that of centralized models, while offering improved scalability and stronger privacy guarantees.

Similarly, other approaches~\cite{Lee_2023, Ciaramella2026_IST} proposed a \gls{fl} approach for converting Android \apks to grayscale using Stream Order imaging techniques. They employ locally trained Convolutional Neural Networks (CNNs), sharing only model parameters with the aggregation server rather than raw application data. Even in these cases, the approach achieved results comparable to those of centralized approaches.

Despite growing interest in federated learning, it is not always the most practical solution for privacy-aware malware detection on Android, as it (i) requires a complex client-server infrastructure to coordinate training and result aggregation across devices; and (ii) introduces operational challenges to device availability, heterogeneous hardware capabilities, and communication overhead. Additionally, if real devices are employed, they (iii) introduce train overhead due to compression of models to be suitable for mobile and limited hardware resources.
For these reasons, \gls{fl} is not always the right choice for protecting user privacy, and simpler, centralized, static-based approaches can be easier to implement. 

\myparagraph{Centralized Approaches}
For the reasons described before, conventional machine learning techniques based on static feature extraction remain widely adopted. Early work, such as Drebin~\cite{DREBIN}, demonstrated that sparse, interpretable feature representations can already achieve strong performance while keeping privacy risks low by extracting features from the application manifest and code. 
Later research explored the extraction of a richer set of features, such as API semantic information~\cite{Zhang_APIGraph} and API graph modeling~\cite{Mariconti_MamaDroid}, while maintaining comparable performance. However, since these approaches employ \gls{ml} algorithms to learn patterns, they all suffer from \emph{train} and \emph{inference} risks, as they do not employ any method to protect against such attacks.

Among conventional Android malware detection, there are those that are signature-based. These typically satisfy all privacy risk requirements by employing heuristics to match known patterns or cryptographic hashes against a database of identified malware samples. However, the performance remains low as they are ineffective against previously unseen malware, including zero-day threats and polymorphic variants~\cite{Khraisat_Gondal_Vamplew_Kamruzzaman_2019, Prasad2025}. As a result, while they remain useful for identifying known malware families, they are generally insufficient for detecting modern Android malware.

Furthermore, dynamic analysis can be characterized as privacy-aware when conducted within isolated, synthetic sandbox or emulated environments rather than on physical user devices. 
The execution of applications in a controlled setting containing no authentic user data, personally identifiable parameters such as IMEI or MAC addresses, or real-world behavioral traces ensures that the detection process operates in strict accordance with the principle of data minimization. Under this paradigm, the benefits of behavioral monitoring are preserved without incurring the privacy risks inherent in real-time observation of physical hardware, ensuring that no sensitive information is accessed, aggregated, or disclosed throughout the analysis.

The remaining concern about sandboxed execution is therefore not privacy but detection effectiveness, since prior work comparing real-device and emulator-based dynamic analysis shows that on-device analysis can outperform its sandboxed counterpart because it is not exposed to the anti-emulation evasion techniques used by modern malware \citep{Alzaylaee2017, Alzaylaee2020}. Recent work narrows this gap: CamoDroid mimics real-device behavior to resist evasion \citep{Faghihi2022}, and other approaches combine static and dynamic signals to improve coverage \citep{daCosta2022}. These results indicate that a privacy-driven preference for sandboxed execution no longer implies a detection cost.

\section{Proposed Privacy-Aware Malware Detection Pipeline}\label{sec:meth}
Previous approaches reveal an issue in Android malware detection: \emph{privacy risks} are not merely a consequence of poor data management, but rather a result of design choices for collecting and processing sensitive data. On the contrary, current privacy-aware approaches relying on anonymization, encryption, or federated algorithms still require access to user-generated or device-linked information.

Our approach, instead of treating privacy as a property to be enforced after data collection to mitigate risks, considers it a design choice that starts with \apk processing and feature extraction. We propose a malware detection pipeline that is \emph{privacy-aware} by design; therefore, overall detection stages do not process user data in the very first place, \eg networking information, and behavioral traces, like in the case of commercial antivirus, limiting the need to apply any privacy-enhancing technology to protect sensitive information.
To this end, \autoref{subsec:meth:req} formalizes the design requirements, while~\autoref{subsec:meth:meth} details the privacy-aware methodology for detecting Android malware.


\begin{figure*}[t]
    \centering
    \includegraphics[width=\linewidth]{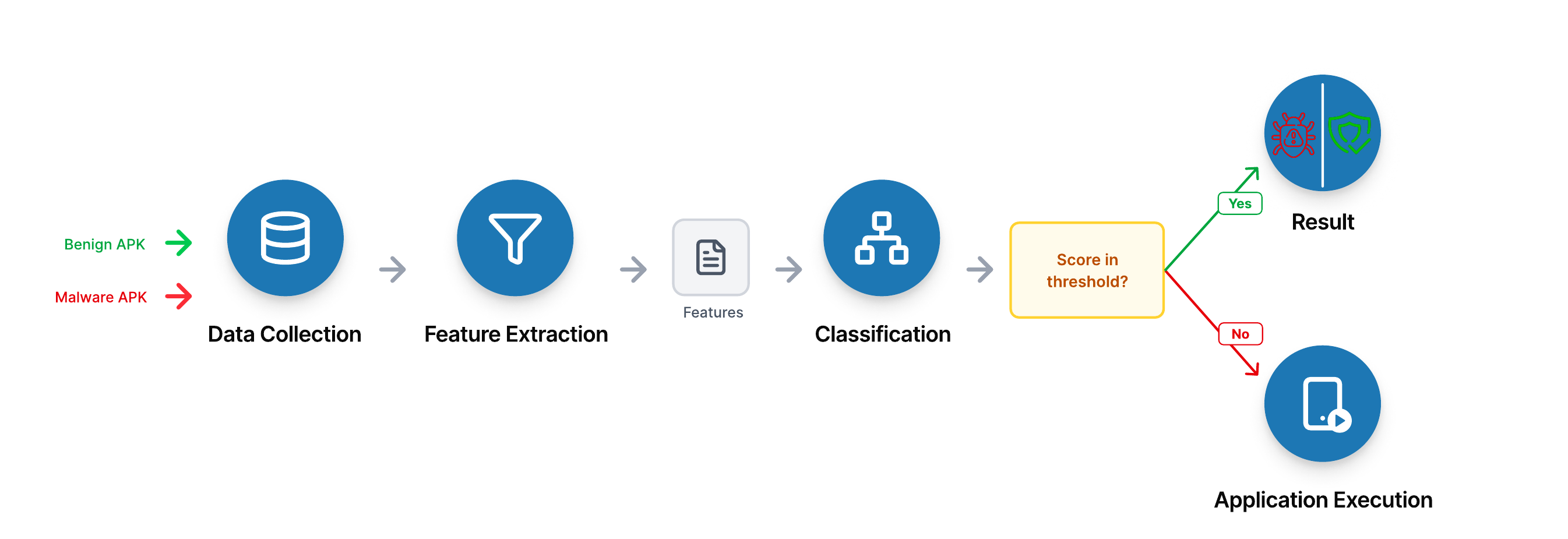}
    \caption{Methodology pipeline consisting in four main stages, \ie Data Collection, Feature Extraction, Classification and Application Execution.}
    \label{fig:pipeline}
\end{figure*}

\subsection{Design Requirements for a Privacy-Aware Pipeline}\label{subsec:meth:req}

In this section, we formalize the design requirements, outlining what should be enforced in a privacy-aware antimalware system. First, privacy constraints, as anticipated, should be integrated directly into the detection methodology rather than added after data collection, thereby limiting the first risk detailed in~\autoref{subsec:background:risks}, \ie the risk of unauthorized access to sensitive data. Indeed, the pipeline should restrict its inputs to what is necessary for detection, excluding \gls{PII}, device identifiers (\eg IMEI and MAC addresses), and behavioral identifiers (\eg app usage, and location information).

Second, static analysis is a preferable choice because it is privacy-aware by design, making it easier to control data processing and feature extraction. However, when static analysis cannot reliably determine whether an application is genuinely benign or potentially malicious, additional inspections become necessary, \ e.g., through dynamic analysis. Indeed, the stealthy nature of modern attacks, together with the widespread adoption of obfuscation and evasion techniques, often prevents static approaches from fully exposing malicious behaviors. As a consequence, more in-depth analyses are required for applications exhibiting suspicious or ambiguous characteristics. In this case, applications must be executed within an isolated sandbox that processes simulated data, \eg network traffic, or location information. 

Third, individual data that are not sensitive in isolation, \eg a single installed application, can enable user profiling when combined with other information such as device identifiers. Therefore, the pipeline must avoid any data aggregation that enables linking single behavioral observations to users or devices.

\subsection{Methodology}\label{subsec:meth:meth}

\autoref{fig:pipeline} shows our methodology, comprising four main stages. First, the \apk is decompiled using static analysis to extract relevant data and characteristics (Data Collection), then we construct a feature vector from the collected information (Feature Extraction) to be ready for the classifier identifying the sample labels (Classification). Eventually, a threshold is applied to the produced scores to determine whether the classifier is confident in its decision, or whether a dynamic analysis is needed to identify malicious patterns (Application Execution). 
In the following, we provide methodological details for each stage.

\myparagraph{(1) Data Collection} 
We first perform static analysis on each \apks in the dataset to extract relevant information from applications. As said before, being privacy-preserving also means having access only to non-sensitive characteristics. For this reason, we adopted Drebin~\cite{DREBIN} as an extractor operating on a rich and interpretable data set derived exclusively from static artifacts. We adopted Drebin because it is widely used in the literature as a baseline model and achieves good performance even against modern malware and its sophisticated techniques, but the presented pipeline can be easily adapted to employ other Android privacy-preserving extractors by design. 
As outlined in~\autoref{subsec:sota:aware}, Drebin inspects mainly manifest files to retrieve (i) Application Components, (ii) Hardware Features, (iii) Permissions, and (iv) Intents, and DEX bytecode to gather APIs, subdivided into (v) Suspicious, and (vi) Restricted calls, (vii) Used Permissions, and (viii) Network Address, \ie IP and URLs, that are hardcoded within the application code. 
All collected data are exclusively taken from inside the \apk with no observation of runtime activity, network traffic, or device-linked identifiers, in line with the data minimization requirement introduced in \autoref{subsec:meth:req}.

\myparagraph{(2) Feature Extraction} 
The data collected in the previous stage cannot be fed directly into a learning model and must first be mapped to a numerical representation. Following the original Drebin~\cite{DREBIN} formulation, we build a global vocabulary $\mathcal{S} = \mathcal{S}_1 \cup \mathcal{S}_2 \cup \dots \cup \mathcal{S}_8$ that collects all distinct data collected across the eight categories on the training partition. 
Each application $x$ is then encoded as a sparse binary vector $\varphi(x) \in \{0,1\}^{|\mathcal{S}|}$, where each entry corresponding to $s \in \mathcal{S}$ is set to $1$ when $s$ appears in the extracted data from $x$ and to $0$ otherwise. $\varphi(x)$ records only the presence of static artifacts contained in the \apk, therefore, it does not encode behavioral patters, or network payloads attributable to a specific user, thereby addressing the feature-extraction risk discussed in \autoref{subsec:background:risks}.

\myparagraph{(3) Classification} 
We then employ a machine learning classifier to identify malicious patterns within the training data. In particular, we employed an SVM model $f$, as proposed by Drebin, which produces a classification score $z = f(\varphi(x))$. 

However, rather than labeling every sample with a binary decision based on a single threshold $\tau$ following~\autoref{eq:single_tau} where $y$ is the chosen label, we employed a confidence-aware classifier with two reject thresholds, $\tau_b$ and $\tau_m$, where $\tau_b$ $<$ $\tau_m$~\cite{threshold_Fumera_Roli_Giacinto_2000, selective_Geifman_El-Yaniv_2017} as shown in~\autoref{eq:multi_tau}; a sample is classified as benign when its score $z$ is at most $\tau_b$, as malicious when its score $z$ is at least $\tau_m$, and is otherwise deferred to the subsequent dynamic analysis stage.

\noindent
\begin{minipage}[t]{0.45\textwidth}
\vspace{0pt}
\[
y =
\begin{cases}
1 & \text{if } z > \tau \\
0 & \text{if } z < \tau
\end{cases}
\tag{1}
\label{eq:single_tau}
\]
\end{minipage}
\\
\begin{minipage}[t]{0.45\textwidth}
\vspace{0pt}
\[
y =
\begin{cases}
1 & \text{if } z > \tau_m \\
0 & \text{if } z < \tau_b \\
? & \text{if } \tau_b < z < \tau_m
\end{cases}
\tag{2}
\label{eq:multi_tau}
\]
\end{minipage}

\vspace{2em}
This formulation reflects the idea that an overconfident static decision based on a single threshold is more harmful than abstention, \ie if there are samples difficult to be correctly classified since their score $z \in I(\tau)$, it is better for the model to be uncertain rather than make a confident but potentially wrong decision. This is especially important in our setting to make confident decisions, reduce false negatives, and prevent the need to gather additional data from the next stage in the pipeline.

Both thresholds $\tau_b$ and $\tau_m$ are selected at runtime on the validation set, after the model has been trained. For each candidate threshold, we compute performance scores and fine-tune them to maintain F1-scores above a fixed admissible floor $\varepsilon_{F1}$ for both benign and malware classes. Among all thresholds satisfying this constraint, we select the most conservative ones, \ie the highest $\tau_m$ and the lowest $\tau_b$, so that the classification stage commits to a decision only when the classifier is confident in the predicted class.
Additionally, we enforce a minimum gap $\Delta_{\tau}$ between the two thresholds to avoid degenerate configurations in which the deferral region collapses, effectively reducing classification at this stage to a single threshold. 
If the F1 constraint cannot be satisfied for either class, the procedure defaults to the thresholds that maximize the per-class F1 while still respecting the gap constraint. Once $\tau_b$ and $\tau_m$ are determined, the model is retrained on the combined training and validation sets and then evaluated on the test set.


\myparagraph{(4) Application Execution} 
In the final stage, samples whose score $z \in (\tau_b, \tau_m)$ are those that the static classifier in stage (3) cannot identify at the chosen confidence level. Therefore, we forward them to a second-level analysis stage based on dynamic techniques.
Consistent with our privacy-aware design, we do not perform such analysis on a real user device, where personal data could be exposed to a centralized server and the associated privacy risks.
Instead, we rely on ANY.RUN \cite{any-run} as a proxy for a privacy-aware dynamic analysis service. ANY.RUN sanboxes execute each submitted application in an isolated environment containing no user content or device-linked artifacts; behavioral signals such as system calls, file system activity, and network endpoints are therefore observed in an environment functionally equivalent to a clean device dedicated solely to analysis. Under this configuration, the visibility required by dynamic inspection is preserved while the data minimization and least-privilege requirements introduced in \autoref{subsec:meth:req} are not violated, since no genuine user data ever enters the analysis loop.

\section{Experimental Evaluation}\label{sec:eval}

In this section, we first outline the evaluation settings, \ie dataset and pipeline implementation, in~\autoref{subsec:eval:set}, and then we discuss the obtained results in~\autoref{subsec:eval:res} to answer the following research questions:

\vspace{0.5em}
\noindent\textbf{RQ1. How effective is Drebin in identifying malware?}

\vspace{0.5em}
\noindent\textbf{RQ2. How effective is a threshold-based methodology?}

\vspace{0.5em}
\noindent\textbf{RQ3. How effective are online privacy-aware sandboxes?}

\vspace{0.5em}
\noindent\textbf{RQ4. How privacy-aware is the methodology?}

\subsection{Experimental Settings}\label{subsec:eval:set}

We describe below the experimental setup used to evaluate the Android malware detection pipeline and its privacy preservation.

\begin{figure*}[t]
    \centering
    \includegraphics[width=0.75\linewidth]{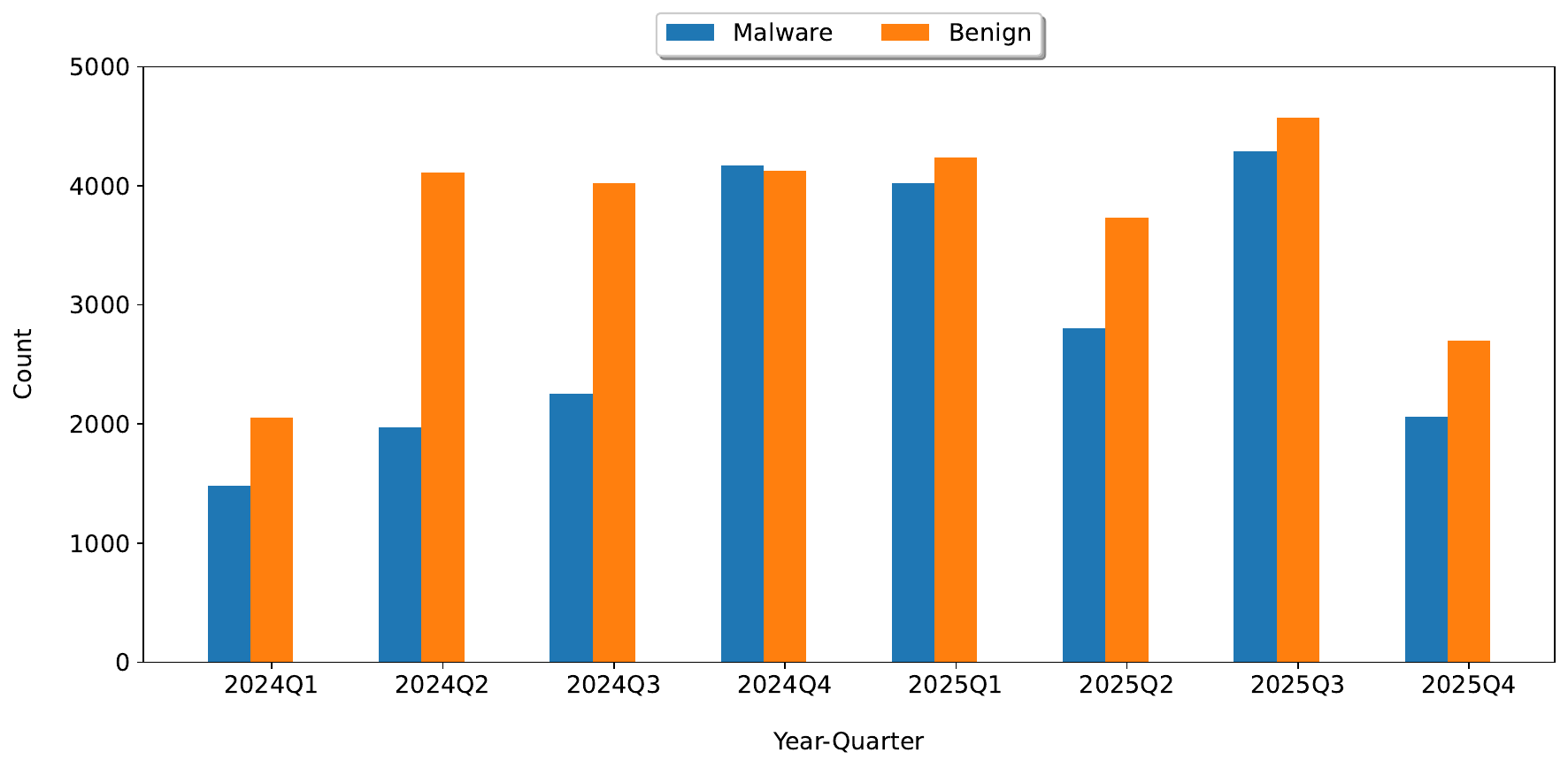}
    \caption{Malware and Benign dataset composition per year and quarter after preprocessing.}
    \label{fig:dataset_quarters}
\end{figure*}

\myparagraph{Dataset}
We conduct experiments employing a dataset of Android applications collected from \gls{vt}. In particular, we used structured queries for \gls{vt} APIs to \emph{(i)} identify malicious and benign samples; \emph{(ii)} infer a development time window to construct a \textit{year-quarter} temporally organized dataset; and \emph{(iii)} reduce label noise from incomplete scans.
As an example, consider the query below:

\begin{verbatim}
    tag:apk AND 
    fs:2025-10-01+ AND fs:2025-12-31- AND 
    ls:2026-02-01+ AND p:0
\end{verbatim}

In this query, the first constraint is handled with the \texttt{p} tag, which indicates the number of \gls{vt} engines that flagged the sample as malicious. We chose zero for benign samples, and greater than five for malicious \apks as recommended by Chow \etal~\cite{Chow2026_Satml}. The \texttt{fs} tag helps fulfill the second constraint, which is necessary to provide temporally consistent experiments~\cite{Pendlebury19_USENIX}. In particular, it specifies the \textit{first-seen} interval to identify the year and quarter in which the sample was submitted to \gls{vt}, which is a valid approach for inferring the time at which a sample "arrives" at the detector~\cite{Chow2026_Satml}. Eventually, the \texttt{ls} tag specifies the \textit{last-seen}, ensuring that samples are scanned at least once. This constraint is necessary to mitigate potential label changes over time, which could otherwise introduce false positives or false negatives in the overall analysis.





\autoref{fig:dataset_quarters} reports the number of samples retained after preprocessing, \ie a step to eliminate samples that are not parsed by the framework, for each quarter and class.


\myparagraph{Setting} 
To evaluate the detection models under a realistic deployment scenario, we adopt a temporal split. All samples from 2024 and the first two quarters of 2025 are used for training, while later quarters are reserved for validation and testing. Specifically, applications first seen in 2025~Q3 are used for validation, and those from 2025~Q4 are used for the final test set. This chronological separation prevents information leakage from future samples. It better reflects the real-world setting in which detection models are trained on past data and deployed on previously unseen applications.

The resulting splits contain 38\,992 samples for training, 8\,861 for validation, and 4\,760 for testing. The training set includes 16\,713 malware and 22\,279 benign applications (malware prevalence $0.429$). The validation and test sets contain 4\,288 and 2\,063 malware samples, respectively, with similar class distributions. This setup allows evaluating the proposed pipeline's ability to generalize to newly observed Android applications.

\myparagraph{Implementation}
To ensure reproducibility, we rely on a publicly available implementation of the \textsc{Drebin} classifier~\citep{android-detectors_2026} for both feature extraction and model training. The hyperparameters of the underlying linear SVM, namely the regularization parameter $C$, the minimum and maximum document frequencies $\mathit{min\_df}$ and $\mathit{max\_df}$, and the maximum vocabulary size, are selected through a grid search. Each configuration is trained on the training set and evaluated on the validation set, and retaining the configuration that maximizes the malicious class $F_1$ score, using recall as a tie-breaker, subject to an overall validation-set accuracy of at least $0.85$. This procedure produces $C = \texttt{0.03}$, $\mathit{min\_df} = \texttt{1}$, and $\mathit{max\_df} = \texttt{0.9}$. The selected configuration is then retrained on the training and validation sets and evaluated again on the test set. For the dual-threshold selection procedure described in \autoref{subsec:meth:meth}, we set an admissible per-class $F_1$ floor $\varepsilon_{F_1} = 0.85$ and a minimum inter-threshold gap $\Delta\tau = 0.10$. The latter controls the minimum acceptable classification quality of samples, while the former prevents the rejection interval from collapsing into a single-threshold configuration.

For ANY.RUN, instead, we used APIs and manual sample submission. Due to sandbox quotas, we do not submit all deferred samples; instead, we use only $50$ samples per class. For each sample, ANY.RUN returns either \textit{Malicious}, or \textit{Suspicious}, or \textit{No Threats Detected}, which we record as the outcome of the dynamic stage.
Each sample is submitted twice, first through the API without interaction due to limited automatic input generation in Android environments, and then manually. This lets us assess whether manual interaction plays a role in exposing malicious behavior during dynamic analysis.

\begin{figure*}[t]
    \centering
    \includegraphics[width=0.8\linewidth]{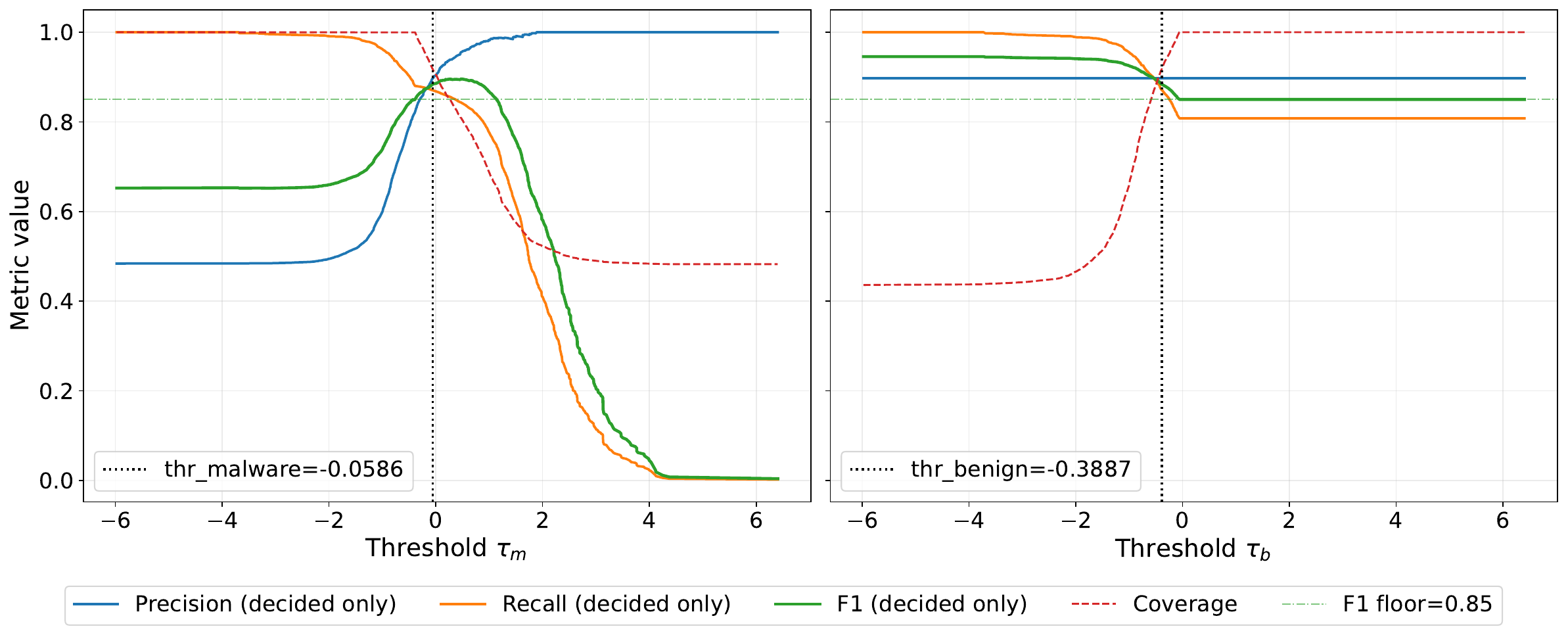}
    \caption{Impact of thresholds on the detection effectiveness measures by Precision, Recall, and F\textsubscript{1} score on samples classified with a strong confidence. On the left are malicious class performances, while on the right the benign class performances.}
    \label{fig:thresholds}
\end{figure*}

\subsection{Experimental Results}\label{subsec:eval:res}


To evaluate our detection pipeline and assess the benefit of the dual-threshold methodology, we compare three settings summarized in~\autoref{tab:classifier_comparison}. In particular, we first employed the original Drebin implementation with a single threshold set to $0.5$ (Setting 1). Then, we optimized the single threshold implementation to achieve the maximum F\textsubscript{1} score (Setting 2), and eventually, we tested the dual-threshold implementation (Setting 3) optimized to reach maximum F\textsubscript{1} score, while also minimizing the number of samples rejected and analyzed in Stage (4) Application Execution.

\myparagraph{Setting 1}
Using the original Drebin hyperparameters with a fixed threshold of $0.5$, the classifier achieves $83.03\%$ test accuracy but with a pronounced precision--recall imbalance: malware precision reaches $0.931$, while recall falls to $0.657$ ($\text{F}_1 = 0.770$). The classifier, therefore, misses roughly one in three malware samples, making a positive prediction only when the score is well above the decision boundary. This confirms that the original threshold is overly conservative for this temporal distribution and motivates re-tuning.

\begin{table}[t]
\caption{Comparison of static classifier configurations on validation and test sets.
Metrics for Setting~3 are computed on the decided subset only (deferred samples excluded).}
\label{tab:classifier_comparison}
\centering
\resizebox{\columnwidth}{!}{
\begin{tblr}{
  colspec = {l l cccc c},
  rows = {rowsep = 3pt, valign = m},
  hline{1,Z} = {solid, wd=1pt},
  hline{2} = {3-6}{solid, leftpos=-0.5, rightpos=-0.5, endpos},
  vline{2-7} = {solid}
}
& & \SetCell[c=4]{c}{\textbf{Metrics}}\\
& \textbf{Exp. Settings} & Acc. & P & R & F\textsubscript{1} & \textbf{Deferred}\\
\hline\hline
\SetCell[r=3]{c}{\rotatebox[origin=c]{90}{\textbf{Val}}}
  & (1)~Original $\tau\!=\!0.5$       & 0.81 & 0.96 & 0.64 & 0.76 & --- \\
  & (2)~Single $\tau\!=\!-0.215$      & 0.86 & 0.87 & 0.85 & 0.86 & --- \\
  & (3)~Dual $\tau_b / \tau_m$         & 0.88 & 0.90 & 0.87 & 0.89 & 8.2\% \\
\midrule
\SetCell[r=3]{c}{\rotatebox[origin=c]{90}{\textbf{Test}}}
  & (1)~Original $\tau\!=\!0.5$       & 0.83 & 0.93 & 0.66 & 0.77 & --- \\
  & (2)~Single $\tau\!=\!-0.215$      & 0.86 & 0.85 & 0.84 & 0.84 & --- \\
  & (3)~Dual $\tau_b / \tau_m$         & 0.89 & 0.88 & 0.87 & 0.87 & 6.7\% \\
\end{tblr}
}
\end{table}

\myparagraph{Setting~2}
Selecting the decision threshold on the validation set with the tuned hyperparameters already yields a substantial correction. The optimized threshold $\tau = -0.215$ raises test accuracy to $86.53\%$ and narrows the precision--recall gap to $0.846$/$0.842$ ($\text{F}_1 = 0.844$), eliminating most of the high-miss-rate problem of Setting~1 at a modest precision cost of $0.085$. 
This result addresses RQ1. Drebin remains an effective static baseline under a realistic temporal split, and threshold selection alone accounts for a significant share of the performance gap between the naive and optimized configurations, remaining privacy-preserving.

\myparagraph{Setting~3}
\autoref{fig:thresholds} shows how the choice of thresholds affects classifier effectiveness for both malicious and benign classes. For each candidate value of $\tau_m$ (left), and $\tau_b$ (right), the plot reports precision, recall, and F\textsubscript{1} score on samples for which the classifier is confident, \ie the fraction of validation samples for which the classifier receives a decision based on class scores represented by coverage in the Figure.

Three results can be derived from~\autoref{fig:thresholds}. First, the higher is $\tau_m$, the more conservative the classifier is when assigning samples to malware classes. This leads to higher precision in malware classification, since fewer benign samples are misclassified as malicious. However, such behavior reduces recall, as a larger number of actual malware is no longer detected. At the same time, the percentage of deferred malware samples decreases, since fewer uncertain samples fall into the deferred region.

Conversely, lowering $\tau_b$ has a positive effect on the benign class, as the classifier improves performance across all metrics, reducing the number of deferred benign samples. However, we cannot reduce it to obtain the best scores on benign for three main reasons. First, $\tau_b$ should be greater than $\tau_m$, as specified in~\autoref{subsec:meth:meth}, and we want to achieve good performance for both classes, especially for those that are more critical. 
Overall, the choice of the thresholds represents a trade-off between classification performance and deferment behavior. The objective is to identify a balanced operating point that yields strong predictive metrics while keeping the percentage of deferred samples reasonably low, without altering the classifier's original effectiveness across both classes.

As shown in~\autoref{tab:detect_comparison}, the selected thresholds achieve a F\textsubscript{1} of $0.89$ in validation, and $0.87$ in the test set with only a minimal fraction of them (8.2\% and 6.7\% respectively) that fall in the rejection interval and are forwarded to the App Execution stage. Compared to Setting 2, the dual-threshold formulation improves all metrics at the cost of losing deferred samples. This experiment addresses RQ2. The classifier's decision in high-confidence regions yields higher-quality static predictions than a single threshold that imposes a forced cut-off.

\myparagraph{Results of ANY.RUN}
\autoref{tab:anyrun} reports the results for Stage (4) on 50 malware and benign samples submitted twice to ANY.RUN sandbox, as specified in~\autoref{subsec:eval:set}, that returns \emph{Malicious}, \emph{Suspicious}, or \emph{No Threats Detected}.

Under automated execution, ANY.RUN detects only 14\% of malware and 6\% of suspicious files, leaving the rest undetected. Instead, under manual interaction, the majority of samples, \ie the 80\% are identified as malicious, or at least suspicious.

Instead, for benign samples submitted via API calls, the results show that the majority of \apks are detected as malicious or suspicious, and only 32\% receive a clear verdict. Manual interaction does not improve this: 40\% are still flagged as malicious and 38\% as suspicious, leaving only 22\% correctly cleared.

This high rate of false positives is not surprising, as these samples are rejected since some static artifacts resemble those of malicious. Additionally, if dynamic execution classifies them as malware, it is plausible that they also rely on suspicious API calls or behavioral IoCs. 
For instance, applications such as custom camera apps, third-party VPNs, or VoIP clients legitimately request access to sensitive APIs and system resources (e.g., network sockets, microphone, or accessibility services) that overlap with those commonly abused by malware, making them difficult to distinguish from legitimate applications.
This points to a limit of automated dynamic analysis on borderline samples like these, and manual inspection would likely be required to settle these cases. 
This addresses RQ3. Application execution can be effective against deferred malware only when a human operator interacts with the application, and its automated verdicts on benign samples in the deferred band are not reliable.

\begin{table}[t]
\caption{ANY.RUN dynamic analysis outcomes for 50 malware and 50 benign samples drawn from the deferred set.
 Samples are submitted twice (without and with manual interaction)}
\label{tab:anyrun}
\centering
\resizebox{\columnwidth}{!}{
\begin{tblr}{
  colspec = {l l ccc},
  rows = {rowsep = 3pt, valign = m},
  hline{1,Z} = {solid, wd=1pt},
  hline{2} = {3-5}{solid, leftpos=-0.5, rightpos=-0.5, endpos},
  vline{2-5} = {solid}
}
& & \SetCell[c=3]{c}{\textbf{Verdict}}\\
& \textbf{Condition} & Malicious & Suspicious & No Threats \\
\hline\hline
\SetCell[r=2]{c}{\textbf{Malware}}
  & No interaction    & 7 (14\%)  & 4 (8\%)   & 39 (78\%) \\
  & With interaction  & 27 (54\%) & 13 (26\%) & 10 (20\%) \\
\midrule
\SetCell[r=2]{c}{\textbf{Benign}}
  & No interaction    & 22 (44\%) & 13 (26\%) & 16 (32\%) \\
  & With interaction  & 20 (40\%) & 19 (38\%) & 11 (22\%) \\
\end{tblr}
}
\end{table}

\subsection{Discussion}\label{subsec:res:discussion}
The proposed pipeline addressed RQ4 by treating privacy as a structural property of the methodology rather than a secondary consideration. By relying on static artifacts extracted directly from the \apk, the pipeline never processes user-generated data or device-linked information, therefore satisfying the data minimization \citep{DataMinimization} and aggregation requirements discussed in \autoref{subsec:meth:req}.
For samples deferred to dynamic analysis, ANY.RUN serves as a concrete example of a sandbox that preserves these properties since applications are executed in an isolated environment containing no authentic user content, so that the behavioral visibility required at this stage does not pose a risk for the user's privacy. Any equivalent sandboxed solution would satisfy the same requirement, and the choice of ANY.RUN should be understood as a placeholder rather than a dependency of the proposed approach.
\section{Conclusion and Future Work}\label{sec:conclusion}
We presented a four-stage Android malware detection pipeline designed to be privacy aware from the ground up. Rather than collecting sensitive data and then applying privacy enhancing techniques to manage the associated risks, our approach avoids touching user data in the first place: if a detection system never accesses sensitive information, there is nothing to anonymize or obfuscate, simplifying not only the analysis as a smaller number of data is extracted, but also the privacy is taken into account by design without any privacy-aware technique. Static features are extracted directly from APK files without observing runtime behavior or device-linked identifiers, and a dual-threshold SVM either classifies samples confidently as malware or benign or defers them to a sandboxed dynamic analysis service operating on synthetic data that does not expose any sensitive data to third parties.

On a temporally split VirusTotal dataset spanning 2024 and 2025, the pipeline achieves 0.89 accuracy and a malware $F_1$ of 0.87 on the decided samples, while deferring only 6.7\% of the test set. The dual-threshold formulation consistently outperforms both the original Drebin configuration and the single-tuned threshold, confirming that abstaining on hard cases is more useful than forcing a decision on samples whose scores sit close to the boundary.
The dynamic stage results are more mixed and arguably the most interesting part of the evaluation. Automated sandbox execution detects very little of the deferred malware, while manual interaction recovers most of it, with 80\% of samples flagged as malicious or suspicious. This gap is large enough to be the most significant takeaway of the evaluation: malware samples that were already uncertain under static analysis are likely those that hide their behavior behind user-triggered code paths, and passive observation alone will not expose them.
This directly motivates two directions for future work. First, automating realistic application interactions for dynamic analysis through UI exploration or simulated user input would allow the deferral stage to operate effectively at scale without requiring human effort, and second, replacing ANY.RUN with a self-hosted sandbox would preserve the same privacy guarantees while removing the dependency on an external service, and the pipeline's modular structure makes this substitution straightforward.

\section*{Acknowledgments}
\begin{sloppypar}
This work has been partially supported by the European Union – NextGenerationEU, Mission 4 Component 1 - CUP F53D23009190001 through the Italian Ministry of University and Research, Project PRIN 2022 PNRR ‘‘PAAM - Privacy Aware Anti Malware’’ grant n. P20225J5YS.
\end{sloppypar}
\bibliographystyle{splncs04}
\bibliography{references}

\end{document}